# *Ab initio* optimization of phonon drag effect for lower-temperature thermoelectric energy conversion

Jiawei Zhou[1], Bolin Liao[1], Bo Qiu[1], Samuel Huberman[1], Keivan Esfarjani[2,3], Mildred S. Dresselhaus[4,5], and Gang Chen[1*]

[1]Department of Mechanical Engineering, Massachusetts Institute of Technology, Cambridge, Massachusetts, 02139, USA

[2]Department of Mechanical and Aerospace Engineering, Rutgers University, Piscataway, New Jersey, 08854, USA

[3]Institute for Advanced Materials, Devices and Nanotechnology, Rutgers University, Piscataway, New Jersey, 08854, USA

[4]Department of Electrical Engineering and Computer Science, Massachusetts Institute of Technology, Cambridge, MA 02139, USA

[5]Department of Physics, Massachusetts Institute of Technology, Cambridge, MA 02139, USA

---

[*] Author to whom correspondence should be addressed. Electronic mail: gchen2@mit.edu. Telephone number: 617-253-3523.

**Abstract**

While the thermoelectric figure of merit zT above 300K has seen significant improvement recently, the progress at lower temperatures has been slow, mainly limited by the relatively low Seebeck coefficient and high thermal conductivity. Here we report, for the first time, success in first-principles computation of the phonon drag effect – a coupling phenomenon between electrons and non-equilibrium phonons – in heavily doped region and its optimization to enhance the Seebeck coefficient while reducing the phonon thermal conductivity by nanostructuring. Our simulation quantitatively identifies the major phonons contributing to the phonon drag, which are spectrally distinct from those carrying heat, and further reveals that while the phonon drag is reduced in heavily-doped samples, a significant contribution to Seebeck coefficient still exists. An ideal phonon filter is proposed to enhance zT of silicon at room temperature by a factor of 20 to ~0.25, and the enhancement can reach 70 times at 100K. This work opens up a new venue towards better thermoelectrics by harnessing non-equilibrium phonons.

**Significance Statement**

It has been well-known that the phonon drag effect – an extra electrical current induced by phonon heat flow via electron-phonon interaction – can lead to unusually high Seebeck coefficient at low temperatures. However, its use for improving thermoelectric performance has been controversial. Here using first principles calculations we examine the phonon drag with detailed mode-specific contributions, and reveal that even in heavily-doped silicon at room temperature, phonon drag can still be significant, which challenges the previous belief that phonon drag vanishes in heavily-doped samples. A phonon filter is designed to spectrally decouple the phonon drag from the heat conduction. Our simulation explores the coupled electron phonon transport and uncovers the possibility of optimizing the phonon drag for better thermoelectrics.

## Introduction

In metals and semiconductors, the lattice vibration, or phonons, induced electron scattering has a major influence on the electronic transport properties(1). The strength of the electron-phonon interaction (EPI) depends on the distribution of electron and phonon populations. The EPI problem was first studied by Bloch(2), who assumed the phonons to be in equilibrium (so-called "Bloch condition") when calculating the scattering rates of electrons caused by EPI, because of the frequent phonon-phonon Umklapp scattering. This assumption is widely adopted for the determination of electronic transport properties at higher temperatures(1, 3), including the electrical conductivity and the normal ("diffusive") Seebeck coefficient. Below the Debye temperature, however, the non-equilibrium phonons become appreciable as the phonon-phonon Umklapp scatterings are largely suppressed, and this assumption becomes questionable. The significance of non-equilibrium phonons on the Seebeck coefficient was first recognized by Gurevich(4). The experimental evidence given later(5, 6) clearly showed an "anomalous" peak of the Seebeck coefficient at around 40K in germanium. To address this unusual observation, Herring proposed that the non-equilibrium phonons can deliver excessive momenta to the electrons via the EPI(7). This process generates an extra electrical current in the same direction as the heat flow, as if the electrons were dragged along by phonons. Therefore this effect has been dubbed "phonon drag"(7), which makes itself distinct from the normal (diffusive) Seebeck effect. Subsequent explorations(8–13) revealed that this effect exists in various material systems. In particular, it has been speculated that phonon drag is responsible for the extremely high Seebeck coefficient $S = -45000 \mu V/K$ experimentally found in $FeSb_2$(14, 15).

The efficiency of thermoelectric materials is characterized by the figure of merit zT, defined as $zT = \sigma S^2 T / \kappa$, where $\sigma, S, \kappa, T$ are the electrical conductivity, Seebeck coefficient, thermal conductivity and absolute temperature, respectively. Thermoelectric energy conversion at lower temperatures (around and below 300K) can benefit a wide range of applications including refrigeration, air conditioning and cryogenic cooling(16) but is also challenging mainly because the diffusive Seebeck coefficient drops in magnitude while the thermal conductivity increases as the temperature decreases. It is therefore tempting to make use of the phonon drag effect to boost the Seebeck coefficient for better thermoelectrics at lower temperatures. However, controversies exist as to whether zT can be enhanced at all by utilizing the phonon drag effect. Theoretical estimations concluded that the phonon drag will not benefit the thermoelectrics(17), while the experiment on silicon nanowires has suggested the possibility of utilizing the phonon drag to reach zT of 1 at 200K(18). One major concern lies in the fact that a significant phonon drag requires phonons to be far away from their equilibrium under a certain temperature gradient (or equivalently, to have long mean free paths), which usually implies a high thermal conductivity. This is reflected in the experimental fact that the low-temperature peak of the phonon drag Seebeck coefficient usually coincides with the peak of the thermal conductivity(19, 20). Although it has been qualitatively known that phonons involved in phonon drag have longer mean free paths than those that carry

heat(7), so far it is not clear to what extent one can decouple the contributions, i.e. keep a significant phonon drag effect while reducing the lattice thermal conductivity, the latter of which has become a common strategy in increasing the thermoelectric efficiency(21–23). Furthermore, it was believed that the phonon drag will become negligible at high doping levels necessary for thermoelectrics(19, 24), because the increased carrier concentration largely weakens the phonon drag (the so-called "saturation effect"(7)).

The key information for a better understanding here is a mode-by-mode quantification of phonon contributions to the phonon drag and the thermal conductivity, individually. An exact description of the phonon drag effect requires the solution of coupled electron-phonon Boltzmann transport equations (BTEs) with accurate matrix elements of various scattering processes (most importantly, the electron-phonon scattering and phonon-phonon scattering). Early theoretical works(1, 5, 7, 25) attempted to concurrently solve the coupled electron-phonon BTEs mostly using a variational approach(1). Although reasonable agreements with experiments were achieved(26, 27), the calculations were limited to the lowest orders of the trial functions for the variational method and assumed simplified scattering models, which lacks predictive power due to the large number of adjustable parameters involved and makes the interpretation of the calculated results less intuitive. An alternative approach to solving the coupled BTEs seeks to partially decouple the electron and phonon transport(28). Compared to the variational method, it has the advantage of a more transparent interpretation of the results in terms of contributions from individual phonon modes. This approach has been applied to Si MOSFETs and GaAs/AlGaAs heterojunctions(29), and to bulk silicon at low carrier concentrations recently by incorporating accurate phonon-phonon scattering rates obtained from first-principles calculation into the model(30). Although the quantitative agreement with experiments at low carrier concentrations was improved, the appearances of adjustable material-dependent parameters, i.e. the deformation potentials, for the crucial electron-phonon scattering processes still limit the predictive power and are not satisfactory for a modern mode-by-mode understanding of the phonon drag effect. In addition, a quantitative evaluation of the phonon drag across a range of carrier concentrations (particularly into the heavily-doped region, which is practically relevant to thermoelectrics) was missing. In this context, a fully first-principles understanding of the phonon drag effect becomes highly desirable. The recent development of first-principles simulation tools(31, 32) has made the mode-specific contributions to the thermal conductivity accessible. Obtaining the same information for phonon drag, however, can be exceedingly challenging due to the ultra-dense sampling meshes entailed by the convergence. In this paper, we undertake the challenge of examining the detailed phonon mode contributions to the phonon drag effect in silicon from first-principles, by combining the recently invented interpolation scheme based on the maximally localized Wannier functions(33, 34) for the determination of EPI matrix elements with an accurate description of the non-equilibrium phonon distributions(31, 32) including first-principles calculations of the intrinsic phonon-phonon interactions as well as the phonon scatterings by electrons. By generalizing the partially decoupled electron-phonon BTE framework(29) into the heavily-doped region, we implemented an *ab initio* computational approach without any adjustable parameters and justified it by comparing to experiments across a wide range of temperatures and carrier concentrations. Our result quantifies the previous explanation of the saturation effect in heavily-doped samples(7) as caused by the increased phonon scattering by electrons. Based on the information revealed, we show that the phonon drag effect can be engineered to enhance the Seebeck coefficient while largely reducing the thermal conductivity by identifying the "preferable" phonon modes and filtering out the others. Following this strategy, an ideal phonon filter is proposed to increase zT in n-type silicon by a factor of 20 to ~0.25 at room temperature, and the enhancement can be as large as 70 times at 100K.

## Results

**Theoretical formalism of phonon drag**

A temperature gradient can concurrently drive a phonon flow and an electron flow, which results in the lattice thermal conduction and the Seebeck effect, respectively. A complete description of the transport requires a full solution of the coupled BTEs for the distribution functions of both electrons and phonons(1):

$$\begin{cases} \mathbf{v}_\alpha(\mathbf{k}) \cdot \frac{\partial f_\alpha(\mathbf{k})}{\partial T} \nabla T - e\mathbf{v}_\alpha(\mathbf{k}) \cdot \frac{\partial f_\alpha(\mathbf{k})}{\partial E} \nabla \varphi = -\frac{f_\alpha(\mathbf{k}) - f_\alpha^0(\mathbf{k})}{\tau_\alpha^*(\mathbf{k})} + \left(\frac{\partial f_\alpha(\mathbf{k})}{\partial t}\right)_{e-ph} \\ \mathbf{v}_\lambda(\mathbf{q}) \cdot \frac{\partial n_\lambda(\mathbf{q})}{\partial T} \nabla T = -\frac{n_\lambda(\mathbf{q}) - n_\lambda^0(\mathbf{q})}{\tau_\lambda^*(\mathbf{q})} + \left(\frac{\partial n_\lambda(\mathbf{q})}{\partial t}\right)_{e-ph} \end{cases} \quad (1)$$

where $\varphi$ is the electrochemical potential, $f$ and n are the distribution functions for electrons and phonons, respectively (the distribution functions with 0 in the superscript denote the equilibrium statistics: Fermi-Dirac for electrons and Bose-Einstein for phonons), and **v** is the group velocity for electrons (with wave vector **k**) and phonons (with wave vector **q**). The electron bands and the phonon branches are denoted using $\alpha(\beta)$ and $\lambda$, respectively. The collision terms on the right-hand-sides of Eq. (1) describe the changes of the distribution functions caused by various scattering mechanisms. We have separated out those due to the electron-phonon coupling (denoted by "e-ph") and assumed that all other scattering processes (electron-impurity, phonon-phonon, phonon-impurity) can be described by the mode-dependent relaxation time $\tau^*$.

The electron-phonon interactions are three-particle processes and involve distribution functions of three different states: the initial electron state $f_{\mathbf{k}\alpha}$, the final electron state $f_{\mathbf{k}'\beta}$ and the participating phonon $n_{\mathbf{q}\lambda}$. Within the linearized Boltzmann equation framework(1), only first-order deviations of the distribution functions from their equilibrium values are kept, which include terms linear in $\Delta f_{\mathbf{k}\alpha} = f_{\mathbf{k}\alpha} - f^0_{\mathbf{k}\alpha}$, $\Delta f_{\mathbf{k}'\beta} = f_{\mathbf{k}'\beta} - f^0_{\mathbf{k}'\beta}$ or $\Delta n_{\mathbf{q}\lambda} = n_{\mathbf{q}\lambda} - n^0_{\mathbf{q}\lambda}$, each characterizing the non-equilibrium state of electrons or phonons. For the normal electrical property calculations(1) (e.g. electrical conductivity and diffusive Seebeck coefficient), the Bloch condition is usually assumed, which implies $\Delta n_{\mathbf{q}\lambda} = 0$ (2). This assumption breaks down when non-equilibrium phonons become appreciable and the appearance of $\Delta n_{\mathbf{q}\lambda}$ in the electron BTE (the first line of Eq. (1)) is essentially responsible for the phonon drag contribution $S_{ph}$ to the Seebeck coefficient. However, the determination of $\Delta n_{\mathbf{q}\lambda}$ in turn requires the knowledge of $\Delta f_{\mathbf{k}\alpha}$, which appears in the phonon BTE (the last term in the second line of Eq. (1)) and makes solving fully-coupled BTEs a formidable task. One further step towards solving Eq. (1) is to realize that the influence of non-equilibrium electrons on the phonon drag effect, indirectly through affecting the non-equilibrium phonons, is a higher-order effect (i.e. $\Delta f_{\mathbf{k}\alpha} \approx 0$ can be assumed in the phonon Boltzmann equation; also see discussions in the supplementary note 2). As a result, the electron-phonon BTEs can be partially decoupled, which leads to a feasible computational approach.

The electron-phonon coupling terms in Eq. (1) also contain parts that characterize electron scattering by equilibrium phonons(3) and phonon scattering by equilibrium electrons(35, 36), which can be separated out and described by additional relaxation times (Methods). The former is usually the dominant scattering mechanism for the electron transport, while the latter weakens the phonon drag at high carrier concentrations, as we will show. We define the total relaxation times $\tau$ taking into account both of these scatterings and also $\tau^*$ using Matthiessen's rule (Methods). Because the phonon BTE is decoupled from the electron BTE, the non-equilibrium phonon distribution can be directly written down as $\Delta n_{\mathbf{q}\lambda} = \tau_{\mathbf{q}\lambda} \mathbf{v}_{\mathbf{q}\lambda} \cdot \nabla T \frac{\partial n_{\mathbf{q}\lambda}^0}{\partial T}$. The electron BTE in Eq. (1) can then be readily solved for the non-equilibrium electron distribution $\Delta f_{\mathbf{k}\alpha}$ given $\Delta n_{\mathbf{q}\lambda}$. Considering the electrical current density given by $\mathbf{j} = e \sum_{\mathbf{k}\alpha} \mathbf{v}_{\mathbf{k}\alpha} \Delta f_{\mathbf{k}\alpha}$, we arrive at the

phonon drag contribution to the Seebeck coefficient besides the diffusive contribution

$$S_{ph} = \frac{2e}{3\sigma\Omega N_{\mathbf{k}} N_{\mathbf{q}} k_B T^2} \sum_{\mathbf{q}\lambda} \left[ \hbar\omega_{\mathbf{q}\lambda} \tau_{\mathbf{q}\lambda} \mathbf{v}_{\mathbf{q}\lambda} \cdot \left( \sum_{\mathbf{k}\alpha,\mathbf{k}'\beta} \left( \tau_{\mathbf{k}\alpha} \mathbf{v}_{\mathbf{k}\alpha} - \tau_{\mathbf{k}'\beta} \mathbf{v}_{\mathbf{k}'\beta} \right) \cdot f^0_{\mathbf{k}\alpha} \left( 1 - f^0_{\mathbf{k}'\beta} \right) n^0_{\mathbf{q}\lambda} \cdot \frac{2\Pi}{\hbar} \right) \right] \quad (2)$$

$$\text{with} \quad \Pi = \pi \left| g_{\alpha\beta\lambda}(\mathbf{k},\mathbf{k}',\mathbf{q}) \right|^2 \cdot \delta\left( E_{\mathbf{k}'\beta} - E_{\mathbf{k}\alpha} - \hbar\omega_{\mathbf{q}\lambda} \right) \cdot \delta\left( \mathbf{k}' - \mathbf{k} - \mathbf{q} \right)$$

where $e$ is the electron charge, $\sigma$ is the electrical conductivity, $\Omega$ is the unit cell volume, $N_{\mathbf{k}}$ and $N_{\mathbf{q}}$ are the number of sampling points in the discrete reciprocal space meshes for electrons and phonons, $E$ is the electron energy, $\omega_{\mathbf{q}\lambda}$ is the phonon frequency, $g_{\alpha\beta\lambda}(\mathbf{k},\mathbf{k}',\mathbf{q})$ is the electron-phonon interaction matrix element (see Eq. (S3) in supplementary note 1) and the two delta functions impose the energy and momentum conservation conditions. The term inside the bracket of Eq. (2) is the phonon drag contribution from each phonon mode. The same formula has been obtained in a previous work under the assumption of weak electron-phonon coupling(28), which suggests that the dominant scattering for electrons is impurity scattering and also the phonon scattering by electrons is neglected. A similar model included the phonon scattering by electrons but was largely simplified(37). Here we have derived Eq. (2) from partially decoupled BTEs without making *a priori* assumptions on the scattering processes, thus extending its applicability particularly to the heavily doped region. We note that a strong coupling between dopants and the host material can alter the band structure at extremely high doping concentrations, where the current formalism will be less accurate without necessary modifications.

**Temperature and carrier concentration dependence**

To justify our numerical implementation (Methods), we first examine the phonon drag Seebeck coefficient in lightly doped silicon. The low doping level implies that the impurity scattering(3) as well as the phonon scattering by electrons(36) plays a negligible role. Similar to the previous work(28, 30), good agreement is obtained between the calculation results and the experimental data(8), from 300K down to 80K for electrons and to 60K for holes (supplementary figure S2). An extremely dense sampling mesh of $100 \times 100 \times 100$ **q**-points in the phonon Brillouin zone is used, which is necessary for the convergence at very low temperatures. We have also noticed that, even though the phonon drag effect is only dominant at very low temperatures, it has influences across a wide range of temperatures, extending beyond the room temperature. This latter fact is consistent with previous simulation work(30, 38) and quenched thermopower experiments(39).

To optimize zT, the carrier concentration is a common experimental variable and usually sits around $10^{19}$ to $10^{21}$ cm$^{-3}$, for achieving a higher electrical conductivity(40). In Fig. 1, the Seebeck coefficient as a function of carrier concentration is shown for the n-type silicon, which will be focused on in the following discussion. At low doping levels, the intrinsic phonon drag Seebeck coefficient should be independent of the carrier concentration(7) and when combined with the diffusive contribution agrees well with the experiment. However, if the phonon drag Seebeck coefficient is assumed to be unchanged, an apparent discrepancy with experiments then occurs above $10^{17}$ cm$^{-3}$ doping concentration. The reduction of the phonon drag effect caused by increased carrier concentration has been known as the saturation effect(7).

As the carrier concentration increases, several scattering processes start to affect the transport properties. Impurity scattering of electrons can largely decrease the mobility but is found from our calculation to have only a small influence on the phonon drag effect (supplementary note 3). It is the scattering of phonons that shortens the phonon mean free paths, therefore leading to a significant reduction of the phonon drag effect. Our calculation confirms that the phonon scattering by electrons(7, 37) is the major reason for this reduction while the phonon-impurity scattering contributes much less. This difference stems from their different phonon frequency dependence of the scattering rates. As will be clear later, phonons that contribute to the phonon drag mostly have low frequencies. For these phonons, the phonon-impurity scattering (scattering rate scales as $\omega^4$)(41) drops much faster with the frequency than the phonon-electron scattering (scattering rate scales as $\omega$)(36) and therefore plays a negligible role in reducing the phonon drag. On the other hand, the phonon-electron scattering drops slower than the intrinsic phonon-phonon scattering (scattering rate scales as $\omega^2$) and will eventually dominate the scattering of low-frequency phonons as the carrier concentration increases (supplementary figure S3). It has been inferred that the saturation effect is associated with the interaction of increased electrons on phonons(7, 9), but the interpretation is often vague due to the use of the variational approach(25) and the lack of quantitative evaluation. By performing first principles simulation, we clarify the reason of the phonon drag reduction as the shortening of phonon mean free paths by electron scatterings, and clearly show that the inclusion of this scattering mechanism leads to a good agreement of the phonon drag with the experiment extending to the heavily doped region (Fig. 1). In spite of this reduction, the phonon drag contribution at high carrier concentrations cannot be simply ignored. As is seen in Fig. 1, at $10^{19}$ cm$^{-3}$ doping concentration, the phonon drag contribution to the Seebeck coefficient is still comparable to the diffusive contribution. This finding is against the previous belief that the phonon drag effect vanishes in heavily-doped samples(19, 24, 38), and makes it possible to optimize the phonon drag in heavily-doped materials for better thermoelectric performances.

**Identification of preferable phonon modes**

Having justified our calculation by looking into the temperature and carrier concentration dependence of the phonon drag, we proceed to quantify how one can benefit from the phonon drag effect for better thermoelectrics. First we examine the phonon mode-specific information in lightly-doped silicon. The accumulated contributions to the phonon drag Seebeck coefficient and to the thermal conductivity from each phonon mode with respect to their frequencies from 100K to 300K are shown in Fig. 2a. Compared to the modes that carry heat, the specific phonons that contribute the most to the phonon drag effect have lower frequencies, indicating that they are closer to the zone center and also possess longer wavelengths (supplementary figure S5), which explains why phonons involved in the phonon drag processes have longer mean free path than those that carry heat (Fig. 2b).

These features were qualitatively understood previously(7, 27, 42) and originate from the energy and momentum conservations in the electron-phonon scattering process, for which the phonon wave vector must be small to connect different electron states close to the band edge with the energies differing by a phonon energy. Here with the full knowledge of the spectral contribution, we can quantitatively determine how important each phonon mode is in contributing to the phonon drag. In order to enhance zT, the factor $S^2/\kappa$ needs to be maximized. Provided that the phonon drag contribution is non-negligible in the total Seebeck coefficient, one can ask whether we can reduce the thermal conductivity without sacrificing the Seebeck coefficient much. According to Fig. 2b, one can achieve this by designing a mean-free-path-selective phonon filter. For example, at 300K phonons with mean free paths shorter than 1 μm contribute around 70% to the total thermal conductivity while contributing negligibly to the phonon drag effect, implying that the thermal conductivity can be reduced by 70% without affecting the Seebeck coefficient much by "filtering out" these phonon modes. At lower temperatures, the accumulated contribution to the phonon drag effect has a larger shift towards the long mean free path region compared with the contribution to the thermal conductivity. Therefore this "decoupling" strategy becomes even more effective at lower temperatures.

Because the phonon drag, as we have revealed, makes a significant contribution even at high carrier concentrations, the "decoupling" strategy naturally extends to the heavily-doped region. The resulting optimal zT achievable can be estimated by selecting those phonons that contribute preferably to the phonon drag than to the thermal conductivity, and then "filtering out" all other phonons as much as possible. We can define a mode-specific figure of merit $\zeta_{\mathbf{q}\lambda}$ as the ratio of the mode's contribution to the phonon drag to its thermal conductivity contribution (Methods). Modes with larger $\zeta_{\mathbf{q}\lambda}$ are desirable to maximize the Seebeck coefficient given an upper bound of the thermal conductivity. Figure 3a shows the largest possible phonon drag Seebeck coefficient with different upper bounds of the thermal conductivity. In heavily-doped silicon, it was previously believed that the phonon drag effect is completely suppressed, especially when one also tries to reduce the thermal conductivity. State-of-the-art material synthesis techniques have shown the capability of reducing the room temperature thermal conductivity of silicon to below 4 W/mK(18, 43) via introducing phonon-blocking nanostructures. Our results show, however, even at such a low value of the thermal conductivity in a $10^{19}$ cm$^{-3}$ doped n-type silicon sample, there still can be a phonon drag contribution that is about 25% of the diffusive contribution, if preferable modes are chosen carefully. Figure 3b compares the zT when selecting preferable modes with that when neglecting the phonon drag effect, assuming that the thermal conductivity is reduced to 4 W/mK. The optimized zT for normal bulk silicon is ~0.01 at 300K around $4\times10^{19}$ cm$^{-3}$ doping concentration(19), which can be boosted to ~0.25 by combining the optimized phonon drag effect and the reduced thermal conductivity. On the other hand, the conventional means of nanostructuring to reduce the thermal conductivity is to introduce nanoscale grain boundaries or precipitates that strongly scatter long wavelength phonons, by which the phonon drag effect will also be largely weakened, diminishing the possible enhancement of zT by half (Fig. 3b). Furthermore, this enhancement increases as the temperature decreases, reaching a value as large as 70 at 100K (supplementary figure S6). This striking result indicates the large potential of exploiting the phonon drag effect for lower-temperature thermoelectrics.

Figure 4a shows the distribution of the preferable phonon modes as a function of wavelength and frequency, where it is clearly seen that more preferable modes typically have longer wavelengths and lower frequencies. While the discussions above set the upper bound for the enhancement of $S^2/\kappa$, simple selective mechanisms can be devised based on either one of the variables. Here we show the possibility of using nanoclusters as impurities to selectively scatter phonons with different frequencies. The impurity scattering is generally stronger for phonons with higher frequencies and therefore serves as a low-pass filter. Besides, in the volume fraction range we have explored, the electrical conductivity is barely affected (supplementary note 5). Figure 4b shows the thermoelectric figure of merit zT with different volume fractions and nanocluster sizes at 300K. A large enhancement of zT can be achieved, due to the fact that the thermal conductivity is largely reduced while the Seebeck coefficient is less affected. An optimal nanocluster size of 1nm in diameter (supplementary note 5) with a volume fraction of 0.2%, is found to enhance zT by a factor of 5, for which a significant portion of the Seebeck coefficient comes from the phonon drag. At lower teperatures, this enhancement becomes more pronounced (supplementary figure S7).

## Discussion

In this work the phonon drag effect is investigated in unprecedented detail from first-principles to study the possibility of enhancing the thermoelectric performance around and below 300 K. Although it is widely believed that the phonon drag effect is only dominant at very low temperatures and will be largely suppressed when the carrier concentration becomes higher, especially after the thermal conductivity is reduced, we clearly show that even for silicon with a $10^{19}$ cm$^{-3}$ carrier concentration at room temperature and a low thermal conductivity of 4 W/mK, the phonon drag contribution to the Seebeck coefficient is still appreciable if preferable phonon modes are carefully selected. If these modes are instead scattered by the nanostructures to reduce the thermal conductivity, the phonon drag will also be destroyed, which leads to a significant drop in the possible enhancement of zT. The benefit of the optimized phonon drag effect becomes greater when the temperature is decreased. Moreover, we have proposed practical phonon filtering mechanisms based on the identification of preferable phonon modes. A phonon frequency filtering approach based on nanocluster scattering is shown to have the potential of enhancing zT significantly due to the combined effect of a reduced thermal conductivity and the optimized phonon drag.

Here we discuss some implications of our results in the context of the previous experimental work. The widely-applied way of using grain boundaries as phonon scatterers(21), though efficient in reducing the thermal conductivity, may also destroy the phonon drag effect in certain materials at low temperatures due to the scattering of long wavelength phonons(15). On the other hand, there have been extensive experiments on SiGe alloys(44, 45), which exploit the high frequency phonon scatterings instead. However, these alloys typically contain large fractions of both Si and Ge, where the perturbation theory framework breaks down and the phonon drag effect needs to be reconsidered. We should also mention that the experiment showing zT of 1 in nanowires(18) cannot be simply explained by our results, because only bulk properties are examined in this work while phonons in their experiment are claimed to have 1D-like features. Nonetheless, our results present a rational principle to maintain the phonon drag effect while reducing the thermal conductivity by the appropriate use of a phonon filter.

The computational formalism we implemented to solve the coupled electron-phonon BTEs for the phonon drag effect should be generally applicable beyond silicon. We envisage that along this path more material systems can be systematically studied for quantitative understandings of their coupled electron-phonon transport. We also noticed recent experimental findings(20) of enhanced Seebeck coefficients of a thin conductive layer deposited on an insulating substrate, and the possible explanation that the phonons generated in the substrate with long mean free paths "leak" through the interface and drag the electrons. Generalizing the current formalism to understanding this interfacial phonon drag phenomenon will provide more space for tuning the phonon drag property of materials. Furthermore, other filtering mechanisms can be designed and engineered to better select the preferable phonon modes, thereby bringing benefits to thermoelectric applications, particularly at lower temperatures.

## Methods

### Relaxation times in Boltzmann equations

The electron-phonon scattering terms in the coupled BTEs can be linearized to only keep terms linearly proportional to the non-equilibrium distribution functions. After neglecting the influences of non-equilibrium electron distributions on the phonon system, it can be shown that the electron-phonon coupling has the following form

$$\begin{cases} \left(\dfrac{\partial f_\alpha(\mathbf{k})}{\partial t}\right)_{e-ph} \simeq -C_{\mathbf{k}\alpha} \cdot \Delta f_{\mathbf{k}\alpha} + F\left(\Delta n_{\mathbf{q}\lambda}\right) \\ \left(\dfrac{\partial n_\lambda(\mathbf{q})}{\partial t}\right)_{e-ph} \simeq -C_{\mathbf{q}\lambda} \cdot \Delta n_{\mathbf{q}\lambda} \end{cases} \quad (3)$$

where $F$ is a linear function of the non-equilibrium phonon distribution. $C_{\mathbf{k}\alpha}$ and $C_{\mathbf{q}\lambda}$ are constants involving the equilibrium electron and phonon distributions (see detailed derivation in supplementary note 1). We define relaxation times due to equilibrium EPI for electrons and phonons respectively: $\tau_\alpha^{eph}(\mathbf{k}) = 1/C_{\mathbf{k}\alpha}$ and $\tau_\lambda^{eph}(\mathbf{q}) = 1/C_{\mathbf{q}\lambda}$. These can be incorporated into Eq. (1), by redefining the total relaxation times. For electrons, the total relaxation time $1/\tau_\alpha(\mathbf{k}) = 1/\tau_\alpha^*(\mathbf{k}) + 1/\tau_\alpha^{eph}(\mathbf{k})$ includes both electron-impurity scatterings and the electron-phonon scattering. For phonons, $1/\tau_\lambda(\mathbf{q}) = 1/\tau_\lambda^*(\mathbf{q}) + \tau_\lambda^{eph}(\mathbf{q})$ contains phonon-impurity scatterings, phonon-phonon scatterings as well as the phonon scattering by electrons.

### Electron phonon interaction

The equilibrium properties of electrons and phonons are calculated using the density functional theory and density functional perturbation theory as implemented in the QUANTUM ESPRESSO package(46). We use the norm-conserving pseudopotential with the Perdew and Zunger(47) local density approximation (LDA) for the exchange-correlation functional and a cutoff energy of 60 Ryd. A $12\times12\times12$ **k**-mesh is used for the electronic band structure and a $6\times6\times6$ **q**-mesh is used for the phonon dispersion. On these coarse meshes we also obtain electronic wavefunctions, phonon mode-specific perturbing potentials as well as the electron-phonon coupling matrix elements. We use the EPW package(48) with the Wannier interpolation scheme(33) to map these information to a much denser mesh, and then carry out the integration over the Brillouin zone to obtain scattering rates due to the electron phonon interaction (these include electron scattering by phonons and phonon scattering by electrons; for electron-impurity scattering see supplementary note 3). More simulation details can be found in supplementary note 3.

**Lattice conduction calculation**

The phonon relaxation time $\tau_{\mathbf{q}\lambda}$ is required for the calculation of both the phonon drag Seebeck coefficient and the thermal conductivity and characterizes the non-equilibrium phonons. Here we briefly summarize the method and more details are available in the literature(31, 32). Because phonon relaxation times are essentially characterized by the anharmonic force constants (third-order force constants in this calculation), a minimal set of anharmonic force constants is first determined using the symmetry of silicon. They are fitted, based on first principles data regarding forces acting on different atoms and their displacements in a large supercell ($2\times2\times2$ conventional unit cells, 64 atoms), with imposed translational and rotational invariances(49). These third-order force constants are then used to calculate the phonon relaxation times due to phonon-phonon scatterings, via Fermi's golden rule. For heavily-doped samples, the electron scattering of phonons becomes important and will further reduce the phonon relaxation time especially for long-wavelength phonons(36). The total phonon relaxation time $\tau_{\mathbf{q}\lambda}$ combines both phonon-phonon scattering and phonon-electron scattering (for phonon-impurity scattering see supplementary note 3), according to Matthiessen's rule.

**Figure of merit for "preferable" phonon modes**

Each phonon mode labeled by wave vector $\mathbf{q}$ and branch number $\lambda$ makes a contribution to phonon drag given by Eq. (2) and to the thermal conductivity given by $\kappa(\mathbf{q},\lambda) = \frac{1}{3\Omega N_{\mathbf{q}}} \mathbf{v}_{\mathbf{q}\lambda}^2 \tau_{\mathbf{q}\lambda} \hbar \omega_{\mathbf{q}\lambda} (\partial n_{\mathbf{q}\lambda}/\partial T)$. The mode-specific figure of merit $\zeta_{\mathbf{q}\lambda}$ is defined as the ratio between them:

$$\zeta_{\mathbf{q}\lambda} = \frac{S_{ph}(\mathbf{q},\lambda)}{\kappa(\mathbf{q},\lambda)} = \frac{2e}{\sigma N_{\mathbf{k}} k_B T^2} \frac{\mathbf{v}_{\mathbf{q}\lambda} n_{\mathbf{q}\lambda}^0 \cdot \sum_{\mathbf{k}\alpha,\mathbf{k}'\beta} \left( \tau_{\mathbf{k}\alpha} \mathbf{v}_{\mathbf{k}\alpha} - \tau_{\mathbf{k}'\beta} \mathbf{v}_{\mathbf{k}'\beta} \right) \cdot f_{\mathbf{k}\alpha}^0 \left(1 - f_{\mathbf{k}'\beta}^0\right) \cdot \frac{2\Pi}{\hbar}}{\mathbf{v}_{\mathbf{q}\lambda}^2 (\partial n_{\mathbf{q}\lambda}/\partial T)} \quad (4)$$

An upper bound for the thermal conductivity can be described as $\sum_{(\mathbf{q},\lambda)\in C} \kappa(\mathbf{q},\lambda) \leq \kappa_{\max}$, where set $C$ denotes the phonon modes that are selected. Given this constraint, the largest possible phonon drag contribution that one can achieve is obtained by selecting those modes that have figures of merit as large as possible, noting that

$$\max\left( \sum_{(\mathbf{q},\lambda)\in C} S_{ph}(\mathbf{q},\lambda) \right) = \max\left( \sum_{(\mathbf{q},\lambda)\in C} \zeta_{\mathbf{q}\lambda} \cdot \kappa(\mathbf{q},\lambda) \right) \quad (5)$$

The figure of merit defined above distinguishes the "preferable" phonon modes that are more significant in phonon drag from those that are less important, and serves as the criterion to select phonons if one seeks to maximize the phonon drag contribution to the Seebeck coefficient.

## Aknowledgements

We thank Jonathan Mendoza for helpful discussions. This article is based upon work supported partially by S$^3$TEC, an Energy Frontier Research Center funded by the U.S. Department of Energy, Office of Basic Energy Sciences, under Award No. DE-SC0001299/DE-FG02-09ER46577 (for understanding the coupled electron-phonon transport), and partially by the Air Force Office of Scientific Research Multidisciplinary Research Program of the University Research Initiative (AFOSR MURI FA9550-10-1-0533) via Ohio State University (for examining the potential of utilizing the phonon drag effect for cryogenic cooling).

**Figure Legends**

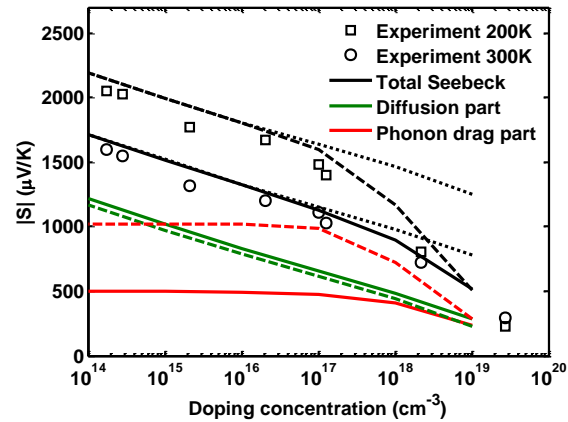

**Figure 1**. Calculated Seebeck coefficient with respect to doping concentrations for n-type silicon at 300K and 200K on a semilog plot. The solid lines describe the calculated results at 300K while dashed lines represent those at 200K. Circles and squares are taken from the experiment(8). At each temperature the total Seebeck coefficient (black) as well as the decomposition into the phonon drag part (red) and diffusion part (green) is shown. Dotted lines are the total Seebeck coefficient calculated using the low doping level value of the phonon drag contribution and assuming this value will not be reduced as the doping concentration increases, i.e. neglecting the "saturation" effect. Compared to the dotted lines, the experimental Seebeck coefficient has a further decrease beyond $10^{17}$ cm$^{-3}$ doping concentration, which is due to the decrease of the phonon drag contribution. This is captured by considering electron scattering of phonons and the resulting Seebeck coefficient (black solid lines for 300K and black dashed lines for 200K) agrees with experiments across the full range of doping concentrations.

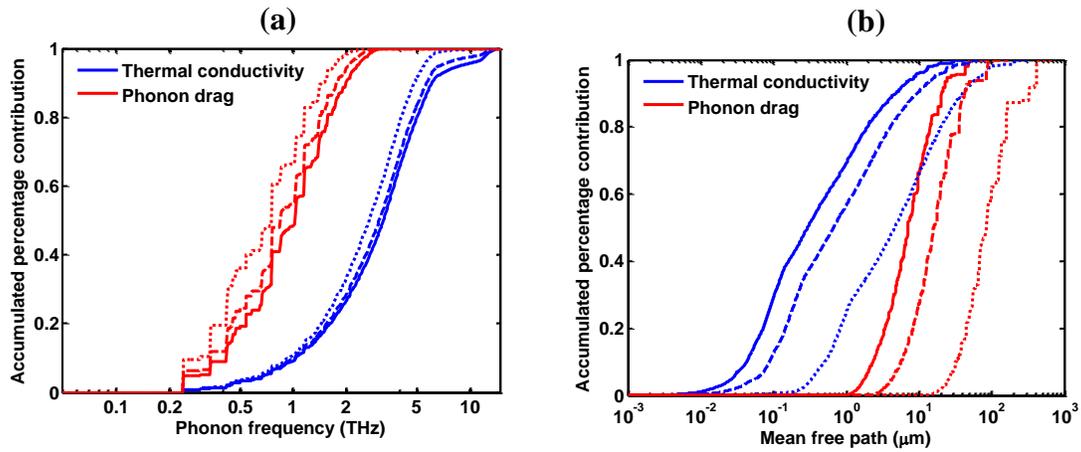

**Figure 2**. Phonon mode-specific accumulated contributions to the phonon drag Seebeck coefficient and the thermal conductivity with respect to (a) phonon frequency and (b) phonon mean free path. Solid lines show the contribution at 300K, while dashed lines are used at 200K and dotted lines at 100K. These results are obtained for lightly-doped n-type silicon.

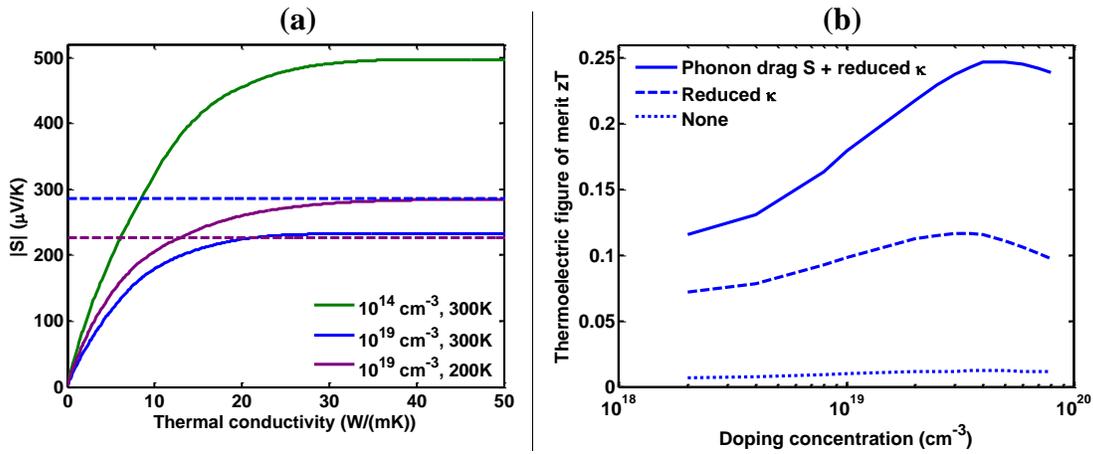

**Figure 3**. (a) Contribution of the most preferable modes to the phonon drag Seebeck coefficient at different reduced thermal conductivity values and (b) the enhancement of zT achieved by selecting preferable modes at 300K for n-type silicon. In (a), for lightly-doped silicon, the thermal conductivity can be reduced to 30 W/(mK) before observing significant diminishment of the phonon drag effect. Dashed lines represent the diffusive Seebeck coefficient for heavily-doped silicon at different temperatures. For heavily-doped silicon, the phonon drag part is still non-negligible and becomes larger compared with the diffusion part when the temperature is decreased. In calculating zT, the experimental data is used for the electrical conductivity as a function of doping concentration for n-type silicon(50).

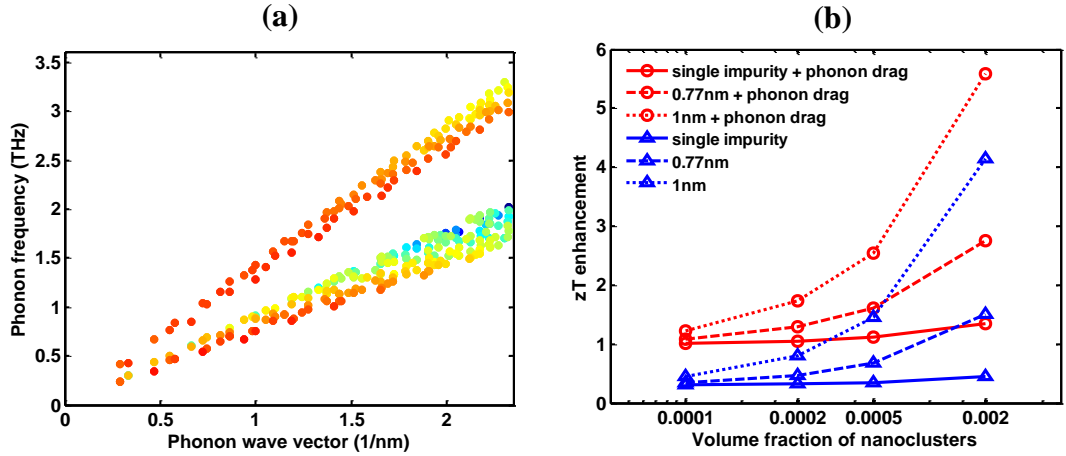

**Figure 4**. (a) Distribution of preferable phonon modes in wave vector and phonon frequency, along with (b) the enhancement of the thermoelectric figure of merit zT compared to that of bulk doped silicon (~0.1 at 300K) with respect to the volume fraction of nanoclusters. Data in part (a) are obtained on a 70x70x70 mesh and only long-wavelength phonons are shown. Red colors represent a higher figure of merit $\zeta_{\mathbf{q}\lambda}$ than blue colors. In part (b), the calculation is done for n-type silicon with a doping concentration of $10^{19}$ cm$^{-3}$ at 300K. Curves labeled with phonon drag include the phonon drag contribution to the Seebeck coefficient, while others negelct the phonon drag effect. The single impurity case is an extreme case of nanocluster scattering, where one nanocluster only contains one impurity atom. For nanoclusters with more than one impurity atom, we calculate the equivalent diameter corresponding to the total volume of unit cells contained in that nanocluster. We note that the 0.77nm size contains 6 unit cells, while the 1nm size contains 14 unit cells. Details of how the nanocluster scattering is treated can be found in supplementary note 6.

**Supplementary Information**

*Ab initio* **optimization of phonon drag effect for lower-temperature thermoelectric energy conversion**

Jiawei Zhou[1], Bolin Liao[1], Bo Qiu[1], Samuel Huberman[1], Keivan Esfarjani[2,3], Mildred S. Dresselhaus[4,5], and Gang Chen[1]

1. Department of Mechanical Engineering, Massachusetts Institute of Technology, Cambridge, Massachusetts, 02139, USA

2. Department of Mechanical and Aerospace Engineering, Rutgers University, Piscataway, New Jersey, 08854, USA

3. Institute for Advanced Materials, Devices and Nanotechnology, Rutgers University, Piscataway, New Jersey, 08854, USA

4. Department of Electrical Engineering and Computer Science, Massachusetts Institute of Technology, Cambridge, MA 02139, USA

5. Department of Physics, Massachusetts Institute of Technology, Cambridge, MA 02139, USA

## Supplementary Note 1. Coupled electron-phonon Boltzmann equation

The analysis based on coupled electron-phonon Boltzmann equation for the phonon drag effect has been given in previous work(1–3). We do not repeat the derivations but provide some necessary details to complete the computational formalism we propose. The coupled electron-phonon Boltzmann equation has been shown in the main text:

$$\begin{cases} \mathbf{v}_\alpha(\mathbf{k}) \cdot \dfrac{\partial f_\alpha(\mathbf{k})}{\partial T} \nabla T - e\mathbf{v}_\alpha(\mathbf{k}) \cdot \dfrac{\partial f_\alpha(\mathbf{k})}{\partial E} \nabla \varphi = -\dfrac{f_\alpha(\mathbf{k}) - f_\alpha^0(\mathbf{k})}{\tau_\alpha^*(\mathbf{k})} + \left(\dfrac{\partial f_\alpha(\mathbf{k})}{\partial t}\right)_{e-ph} \\ \mathbf{v}_\lambda(\mathbf{q}) \cdot \dfrac{\partial n_\lambda(\mathbf{q})}{\partial T} \nabla T = -\dfrac{n_\lambda(\mathbf{q}) - n_\lambda^0(\mathbf{q})}{\tau_\lambda^*(\mathbf{q})} + \left(\dfrac{\partial n_\lambda(\mathbf{q})}{\partial t}\right)_{e-ph} \end{cases} \quad (S1)$$

where $f$ and $N$ represent the distribution functions for electrons and phonons respectively, with equilibrium state described by

$$\begin{cases} f_\alpha^0(\mathbf{k}) = \dfrac{1}{e^{(E_\mathbf{k}^\alpha - \mu)/k_B T} + 1} \\ N_\lambda^0(\mathbf{q}) = \dfrac{1}{e^{\hbar \omega_\mathbf{q}^\lambda / k_B T} - 1} \end{cases} \quad (S2)$$

For electrons, the external driving forces include the electrochemical potential gradient $\nabla \varphi$ and temperature gradient $\nabla T$, while for phonons the only external driving force comes from the temperature gradient. The right-hand-side of Eq. (S1) describes the various scattering events experienced by electrons and phonons.

In equation (S1), we assume that all other scattering mechanisms except the electron-phonon interaction can be described by the mode-dependent relaxation time model, adding these contributions together according to Matthiessen's rule. For electrons $\tau_\alpha^*(\mathbf{k})$ describes the electron-impurity scattering. For phonons $\tau_\lambda^*(\mathbf{q})$ includes both phonon-phonon scattering and phonon-impurity scattering. The scattering rates due to the electron-phonon interaction can be obtained if we have the knowledge of the interaction matrix elements:

$$\left\{\begin{aligned}\left(\frac{\partial f_\alpha(\mathbf{k})}{\partial t}\right)_{e-ph} &= \frac{2\pi}{\hbar}\sum_{\mathbf{k}'\beta,\mathbf{q}\lambda}\left\{\begin{aligned}&-|g_{\alpha\beta\lambda}(\mathbf{k},\mathbf{k}',\mathbf{q})|^2\, n_{\mathbf{q}\lambda} f_{\mathbf{k}\alpha}(1-f_{\mathbf{k}'\beta})\delta(\mathbf{k}'-\mathbf{k}-\mathbf{q})\delta(E_{\mathbf{k}'\beta}-E_{\mathbf{k}\alpha}-\hbar\omega_{\mathbf{q}\lambda})\\ &-|g_{\alpha\beta\lambda}(\mathbf{k},\mathbf{k}',\mathbf{q})|^2\,(n_{\mathbf{q}\lambda}+1) f_{\mathbf{k}\alpha}(1-f_{\mathbf{k}'\beta})\delta(\mathbf{k}'-\mathbf{k}+\mathbf{q})\delta(E_{\mathbf{k}'\beta}-E_{\mathbf{k}\alpha}+\hbar\omega_{\mathbf{q}\lambda})\\ &+|g_{\alpha\beta\lambda}(\mathbf{k}',\mathbf{k},\mathbf{q})|^2\, n_{\mathbf{q}\lambda}(1-f_{\mathbf{k}\alpha}) f_{\mathbf{k}'\beta}\delta(\mathbf{k}-\mathbf{k}'-\mathbf{q})\delta(E_{\mathbf{k}\alpha}-E_{\mathbf{k}'\beta}-\hbar\omega_{\mathbf{q}\lambda})\\ &+|g_{\alpha\beta\lambda}(\mathbf{k}',\mathbf{k},\mathbf{q})|^2\,(n_{\mathbf{q}\lambda}+1)(1-f_{\mathbf{k}\alpha}) f_{\mathbf{k}'\beta}\delta(\mathbf{k}-\mathbf{k}'+\mathbf{q})\delta(E_{\mathbf{k}\alpha}-E_{\mathbf{k}'\beta}+\hbar\omega_{\mathbf{q}\lambda})\end{aligned}\right\}\\ \left(\frac{\partial n_\lambda(\mathbf{q})}{\partial t}\right)_{e-ph} &= \frac{2\pi}{\hbar}\frac{1}{2}\sum_{\mathbf{k}\alpha,\mathbf{k}'\beta}\left\{\begin{aligned}&-|g_{\alpha\beta\lambda}(\mathbf{k},\mathbf{k}',\mathbf{q})|^2\, n_{\mathbf{q}\lambda} f_{\mathbf{k}\alpha}(1-f_{\mathbf{k}'\beta})\delta(\mathbf{k}'-\mathbf{k}-\mathbf{q})\delta(E_{\mathbf{k}'\beta}-E_{\mathbf{k}\alpha}-\hbar\omega_{\mathbf{q}\lambda})\\ &+|g_{\alpha\beta\lambda}(\mathbf{k},\mathbf{k}',\mathbf{q})|^2\,(n_{\mathbf{q}\lambda}+1) f_{\mathbf{k}\alpha}(1-f_{\mathbf{k}'\beta})\delta(\mathbf{k}'-\mathbf{k}+\mathbf{q})\delta(E_{\mathbf{k}'\beta}-E_{\mathbf{k}\alpha}+\hbar\omega_{\mathbf{q}\lambda})\\ &-|g_{\alpha\beta\lambda}(\mathbf{k}',\mathbf{k},\mathbf{q})|^2\, n_{\mathbf{q}\lambda}(1-f_{\mathbf{k}\alpha}) f_{\mathbf{k}'\beta}\delta(\mathbf{k}-\mathbf{k}'-\mathbf{q})\delta(E_{\mathbf{k}\alpha}-E_{\mathbf{k}'\beta}-\hbar\omega_{\mathbf{q}\lambda})\\ &+|g_{\alpha\beta\lambda}(\mathbf{k}',\mathbf{k},\mathbf{q})|^2\,(n_{\mathbf{q}\lambda}+1)(1-f_{\mathbf{k}\alpha}) f_{\mathbf{k}'\beta}\delta(\mathbf{k}-\mathbf{k}'+\mathbf{q})\delta(E_{\mathbf{k}\alpha}-E_{\mathbf{k}'\beta}+\hbar\omega_{\mathbf{q}\lambda})\end{aligned}\right\}\end{aligned}\right.$$

with the electron-phonon interaction matrix element

$$g_{\alpha\beta\lambda}(\mathbf{k},\mathbf{k}',\mathbf{q}) = \left(\frac{\hbar}{2m_0\omega_{\mathbf{q}\lambda}}\right)^{1/2}\cdot\langle\mathbf{k}'\beta|\partial_{\mathbf{q}\lambda}V|\mathbf{k}\alpha\rangle \qquad (S3)$$

In equation (S3), $E_{\mathbf{k}\alpha}$ and $E_{\mathbf{k}'\beta}$ describes the energy of electron states, $\omega_{\mathbf{q}\lambda}$ describes the phonon frequency and the two delta functions impose momentum and energy conservation respectively. In the formula for the electron-phonon interaction matrix element, $m_0$ is the mass of one unit cell, $|\mathbf{k}\alpha\rangle$ and $|\mathbf{k}'\beta\rangle$ describe the eigenstates of electrons, while $\partial_{\mathbf{q}\lambda}V$ is the perturbing potential due to the ionic displacement corresponding to a phonon with wave vector $\mathbf{q}$ and branch number $\lambda$. Note that for the scattering rate for phonons there is an extra $1/2$. This is because when $\mathbf{k}$ and $\mathbf{k}'$ go over the Brillouin zone each $\mathbf{k}_1 \to \mathbf{k}_2$ process is counted twice (let $\mathbf{k}=\mathbf{k}_1$ and $\mathbf{k}'=\mathbf{k}_2$ or vice versa).

In equilibrium, all distribution functions take their equilibrium values. In this case, the scattering rates as shown in (S3) should vanish, because otherwise the state of the system will move away from the equilibrium (see Eq. (S1)). The lowest order approximation for Eq. (S3) comes by taking the first order deviation of the distribution functions, which gives rise to the widely-used linearized Boltzmann equation:

$$\begin{cases} \left(\dfrac{\partial f_\alpha(\mathbf{k})}{\partial t}\right)_{e-ph} \simeq -\left[\sum_{\mathbf{k}'\beta,\mathbf{q}\lambda} F_{\mathbf{k}\alpha}(\mathbf{k}'\beta,\mathbf{q}\lambda)\right]\cdot \Delta f_{\mathbf{k}\alpha} + \sum_{\mathbf{k}'\beta,\mathbf{q}\lambda}\left[F_{\mathbf{k}'\beta}(\mathbf{k}\alpha,\mathbf{q}\lambda)\cdot \Delta f_{\mathbf{k}'\beta}\right] + \sum_{\mathbf{k}'\beta,\mathbf{q}\lambda}\left[F_{\mathbf{q}\lambda}(\mathbf{k}\alpha,\mathbf{k}'\beta)\cdot \Delta n_{\mathbf{q}\lambda}\right] \\ \left(\dfrac{\partial n_\lambda(\mathbf{q})}{\partial t}\right)_{e-ph} \simeq \sum_{\mathbf{k}\alpha,\mathbf{k}'\beta}\left[G_{\mathbf{k}\alpha}(\mathbf{k}'\beta,\mathbf{q}\lambda)\cdot \Delta f_{\mathbf{k}\alpha} + G_{\mathbf{k}'\beta}(\mathbf{k}\alpha,\mathbf{q}\lambda)\cdot \Delta f_{\mathbf{k}'\beta}\right] - \left[\sum_{\mathbf{k}\alpha,\mathbf{k}'\beta} G_{\mathbf{q}\lambda}(\mathbf{k}\alpha,\mathbf{k}'\beta)\right]\cdot \Delta n_{\mathbf{q}\lambda} \end{cases}$$

(S4)

where the coefficients $F$ and $G$ only depends on the equilibrium distribution functions:

$$\begin{cases} F_{\mathbf{k}\alpha}(\mathbf{k}'\beta,\mathbf{q}\lambda) = \left[\left(n^0_{\mathbf{q}\lambda}+f^0_{\mathbf{k}'\beta}\right)\Pi_- + \left(n^0_{\mathbf{q}\lambda}+1-f^0_{\mathbf{k}'\beta}\right)\Pi_+\right] \\ F_{\mathbf{k}'\beta}(\mathbf{k}\alpha,\mathbf{q}\lambda) = \left[\left(n^0_{\mathbf{q}\lambda}+1-f^0_{\mathbf{k}\alpha}\right)\Pi_- + \left(n^0_{\mathbf{q}\lambda}+f^0_{\mathbf{k}\alpha}\right)\Pi_+\right] \\ F_{\mathbf{q}\lambda}(\mathbf{k}\alpha,\mathbf{k}'\beta) = \left[\left(f^0_{\mathbf{k}'\beta}-f^0_{\mathbf{k}\alpha}\right)\Pi_- + \left(f^0_{\mathbf{k}'\beta}-f^0_{\mathbf{k}\alpha}\right)\Pi_+\right] \\ G_{\mathbf{k}\alpha}(\mathbf{k}'\beta,\mathbf{q}\lambda) = \left[-\left(n^0_{\mathbf{q}\lambda}+f^0_{\mathbf{k}'\beta}\right)\Pi_- + \left(n^0_{\mathbf{q}\lambda}+1-f^0_{\mathbf{k}'\beta}\right)\Pi_+\right] \\ G_{\mathbf{k}'\beta}(\mathbf{k}\alpha,\mathbf{q}\lambda) = \left[\left(n^0_{\mathbf{q}\lambda}+1-f^0_{\mathbf{k}\alpha}\right)\Pi_- - \left(n^0_{\mathbf{q}\lambda}+f^0_{\mathbf{k}\alpha}\right)\Pi_+\right] \\ G_{\mathbf{q}\lambda}(\mathbf{k}\alpha,\mathbf{k}'\beta) = \left[\left(f^0_{\mathbf{k}'\beta}-f^0_{\mathbf{k}\alpha}\right)\Pi_- - \left(f^0_{\mathbf{k}'\beta}-f^0_{\mathbf{k}\alpha}\right)\Pi_+\right] \end{cases}$$

(S5)

with

$$\begin{cases} \Pi_- = \dfrac{2\pi}{\hbar}\left|g_{\alpha\beta\lambda}(\mathbf{k},\mathbf{k}',\mathbf{q})\right|^2 \cdot \delta\left(E_{\mathbf{k}'\beta}-E_{\mathbf{k}\alpha}-\hbar\omega_{\mathbf{q}\lambda}\right)\cdot \delta(\mathbf{k}'-\mathbf{k}-\mathbf{q}) \\ \Pi_+ = \dfrac{2\pi}{\hbar}\left|g_{\alpha\beta\lambda}(\mathbf{k},\mathbf{k}',\mathbf{q})\right|^2 \cdot \delta\left(E_{\mathbf{k}'\beta}-E_{\mathbf{k}\alpha}+\hbar\omega_{\mathbf{q}\lambda}\right)\cdot \delta(\mathbf{k}'-\mathbf{k}+\mathbf{q}) \end{cases}$$

denoting processes due to the absorption of a phonon $\omega_{\mathbf{q}\lambda}$ and the emission of a phonon, respectively. The first-order deviations $\Delta f_{\mathbf{k}\alpha} = f_{\mathbf{k}\alpha} - f^0_{\mathbf{k}\alpha}$, $\Delta f_{\mathbf{k}'\beta} = f_{\mathbf{k}'\beta} - f^0_{\mathbf{k}'\beta}$ and $\Delta n_{\mathbf{q}\lambda} = n_{\mathbf{q}\lambda} - n^0_{\mathbf{q}\lambda}$ characterize the non-equilibrium state of electrons and phonons. For normal electrical property calculations, the relaxation time approximation is often used. This approximation naturally arises if we assume that only the distribution of the initial state of the electron deviates from the equilibrium ( $\Delta f_{\mathbf{k}\alpha} \neq 0$ ) and that of the final electron state and of phonons remain at equilibrium ( $\Delta f_{\mathbf{k}'\beta}=0, \Delta n_{\mathbf{q}\lambda}=0$ ), which is essentially the Bloch condition. In this case, the prefactor before $\Delta f_{\mathbf{k}\alpha}$ can be defined as $1/\tau_\alpha^{e-ph}(\mathbf{k})$. The relaxation time $\tau_\alpha^{e-ph}(\mathbf{k})$ is what is usually called the electron-phonon relaxation time (for electrons), which determines the intrinsic mobility of one material. We should also note that $\Delta f_{\mathbf{k}'\beta}$ is essentially neglected because the terms containing $\Delta f_{\mathbf{k}'\beta}$ sum up to approximately zero. In metals and for elastic scattering with impurities, this approximation is not valid and therefore an extra correction term $(1-\cos\theta)$ is often added to the electron-phonon relaxation time, which is called the momentum relaxation time(4). In semiconductors, however, it is proved, based on deformation potential models, that for nearly isotropic scattering, the neglect of $\Delta f_{\mathbf{k}'\beta}$ will not cause much difference(4). It has also been shown(5, 6) that without considering $\Delta f_{\mathbf{k}'\beta}$, good agreement for the electrical properties in silicon with experiments can be achieved, justifying the approximation that terms containing $\Delta f_{\mathbf{k}'\beta}$ can be neglected.

A more important perturbation term from equilibrium comes in the evaluation of the assumption $\Delta n_{\mathbf{q}\lambda} = 0$. It is clear that this assumption makes non-equilibrium phonons have no effect on the electron system. When phonons are far away from equilibrium, assuming $\Delta n_{\mathbf{q}\lambda}$ to be zero is no longer valid. These non-equilibrium phonons described by non-zero $\Delta n_{\mathbf{q}\lambda}$ in the electron system (the last term in the first equation of (S4)) are responsible for the phonon drag effect.

The above picture is based on the Seebeck effect, where a temperature gradient induces a phonon heat flow, which delivers part of its momenta to the electron system and gives rise to an extra current. Because of the Kelvin relation $\Pi = TS$, an extra contribution to the Seebeck coefficient also implies an extra Peltier coefficient. We want to further clarify the phonon drag effect in the context of the Peltier effect, which completes the picture and provides a straightforward derivation of equation (2) in the main text. For the Peltier effect, electrons are first driven by the electric field. These non-equilibrium electrons can then deliver their momenta to the phonon system through the first two terms of the right-hand-side in the second equation of (S4), which can be analogously called the "electron drag" effect but is essentially just the manifestation of the "phonon drag" effect in the Peltier effect. Therefore the phonon system acquires an extra heat flow due to electron motions, which contributes to the Peltier coefficient.

Now we still have one term left (the last term in the second equation of (S4)), which describes the scattering of phonons by equilibrium electrons. The prefactor of $\Delta n_{\mathbf{q}\lambda}$ in the phonon Boltzmann equation can be readily written as $1/\tau_\lambda^{e-ph}(\mathbf{q})$ (just as the definition of electron-phonon relaxation time for electrons). Apparently, higher doping concentrations lead to stronger scattering. We show in our paper that this higher doping concentration is responsible for the reduction of the phonon drag effect in heavily-doped samples compared to lightly-doped samples. Besides, this "electron drag" also accounts for some fraction of the reduction of the thermal conductivity in heavily-doped materials(7).

Having discussed the meaning of each term in equation (S1) and (S4), now we want to make the inclusion of these scattering terms more compact by rearranging them. If we incorporate the first term of the right-hand-side in the first line of Eq. (S4) and the last term in the second line of Eq. (S4) into the relaxation times we have in (S1), the coupled Boltzmann equation now becomes

$$\begin{cases} \left(\dfrac{\partial f_\alpha(\mathbf{k})}{\partial t}\right)_{drift} = -\dfrac{f_\alpha(\mathbf{k}) - f_\alpha^0(\mathbf{k})}{\tau_\alpha(\mathbf{k})} + \sum_{\mathbf{k}'\beta,\mathbf{q}\lambda}\left[F_{\mathbf{k}'\beta}(\mathbf{k}\alpha,\mathbf{q}\lambda)\cdot\Delta f_{\mathbf{k}'\beta}\right] + \sum_{\mathbf{k}'\beta,\mathbf{q}\lambda}\left[F_{\mathbf{q}\lambda}(\mathbf{k}\alpha,\mathbf{k}'\beta)\cdot\Delta n_{\mathbf{q}\lambda}\right] \\ \left(\dfrac{\partial n_\lambda(\mathbf{q})}{\partial t}\right)_{drift} = -\dfrac{n_\lambda(\mathbf{q}) - n_\lambda^0(\mathbf{q})}{\tau_\lambda(\mathbf{q})} + \sum_{\mathbf{k}\alpha,\mathbf{k}'\beta}\left[G_{\mathbf{k}\alpha}(\mathbf{k}'\beta,\mathbf{q}\lambda)\cdot\Delta f_{\mathbf{k}\alpha} + G_{\mathbf{k}'\beta}(\mathbf{k}\alpha,\mathbf{q}\lambda)\cdot\Delta f_{\mathbf{k}'\beta}\right] \end{cases}$$

(S6)

where new relaxation times include part of the electron-phonon coupling

$$\begin{cases} \dfrac{1}{\tau_\alpha(\mathbf{k})} = \dfrac{1}{\tau_\alpha^*(\mathbf{k})} + \sum_{\mathbf{k}'\beta,\mathbf{q}\lambda} F_{\mathbf{k}\alpha}(\mathbf{k}'\beta,\mathbf{q}\lambda) \\ \dfrac{1}{\tau_\lambda(\mathbf{q})} = \dfrac{1}{\tau_\lambda^*(\mathbf{q})} + \sum_{\mathbf{k}\alpha,\mathbf{k}'\beta} G_{\mathbf{q}\lambda}(\mathbf{k}\alpha,\mathbf{k}'\beta) \end{cases}$$

(S7)

and the remaining terms describe the coupling between non-equilibrium states in electron and phonon systems.

## Supplementary Note 2. Derivation of the phonon drag Seebeck coefficient

We have above derived the coupled electron-phonon Boltzmann equation in a compact form as shown in Eq. (S6). The electron relaxation time $\tau_\alpha(\mathbf{k})$ incorporates electron-impurity scattering and electron scattering by equilibrium phonons, while the phonon relaxation time $\tau_\lambda(\mathbf{q})$ contains phonon-phonon interaction, phonon-impurity scattering and phonon scattering by equilibrium electrons. The coupling through the non-equilibrium distribution is manifested by the collision terms that are not described by the relaxation times in Eq. (S6) and those collision terms are responsible for the phonon drag effect. In the main text we have derived the formalism based on the Seebeck picture (the temperature gradient generates a voltage difference), here we adopt the Peltier picture (the isothermal electric field produces a heat flow), which directly provides the phonon drag contribution from each phonon mode and also shows as an explicit proof of the Kelvin relation for the phonon drag effect.

For the Peltier effect, a non-equilibrium distribution of electrons is generated first by the electric field, which will then drive the phonons away from equilibrium. The induced non-equilibrium phonons will now perturb the electrons in a second-order effect, which can be justified by the fact that the phonon drag phenomenon is found to have a small influence on the electrical conductivity(3). Therefore for the electron system, we can then assume that phonons are at equilibrium ($\Delta n_{\mathbf{q}\lambda} = 0$ in the first line in (S6), and note that for the Peltier effect the phonon drag comes in through the last term in the second equation of (S6) instead of the last term in the first equation). This assumption exactly corresponds to the treatment we did in the main text to decouple the phonon transport from the electron system (i.e., assume $\Delta f_{\mathbf{k}\alpha} = 0$ in the phonon BTE) when deriving the phonon drag formula in the Seebeck picture. As a result, the electron distribution function can be directly written down using the relaxation time model $\Delta f_{\mathbf{k}\alpha} = e\tau_{\mathbf{k}\alpha} \mathbf{v}_{\mathbf{k}\alpha} \cdot \nabla\varphi \dfrac{\partial f^0_{\mathbf{k}\alpha}}{\partial E}$ (Here we also assume that the term $\sum_{\mathbf{k}'\beta,\mathbf{q}\lambda}\left[F_{\mathbf{k}'\beta}(\mathbf{k}\alpha,\mathbf{q}\lambda)\cdot\Delta f_{\mathbf{k}'\beta}\right]$ will vanish, which is a commonly used approximation(4)).

The normal electrical conductivity and Peltier coefficient (related to the normal Seebeck coefficient via the Kelvin relation $\Pi = TS$) can be obtained by looking at the charge current and energy current induced by such a non-equilibrium electron distribution, respectively:

$$\begin{cases} \sigma = \left(\frac{e}{3N_{\mathbf{k}}}\sum_{\mathbf{k}\alpha}v_{x,\mathbf{k}\alpha}\Delta f_{\mathbf{k}\alpha}\right)/(-\nabla\varphi) = \frac{e^2}{3N_{\mathbf{k}}}\sum_{\mathbf{k}\alpha}\mathbf{v}_{\mathbf{k}\alpha}^2 \tau_{\mathbf{k}\alpha}\left(-\frac{\partial f_{\mathbf{k}\alpha}^0}{\partial E}\right) \\ \Pi = \left(\frac{e}{3N_{\mathbf{k}}}\sum_{\mathbf{k}\alpha}(E-\mu)v_{x,\mathbf{k}\alpha}\Delta f_{\mathbf{k}\alpha}\right)/\left(\frac{e}{3N_{\mathbf{k}}}\sum_{\mathbf{k}\alpha}v_{x,\mathbf{k}\alpha}\Delta f_{\mathbf{k}\alpha}\right) \\ \qquad = \frac{e}{3\sigma N_{\mathbf{k}}}\sum_{\mathbf{k}\alpha}(E-\mu)\mathbf{v}_{\mathbf{k}\alpha}^2 \tau_{\mathbf{k}\alpha}\left(-\frac{\partial f_{\mathbf{k}\alpha}^0}{\partial E}\right) \end{cases} \quad (S8)$$

The energy current not only can come from the electrons, but also has its origin in the phonon heat flow. As we have discussed, in the isothermal condition the phonons acquire the momentum via the electron-phonon coupling shown by the last term in the second equation of (S6) and lead to an extra heat flow, which manifests the "phonon drag" effect. Given the electron distribution in (S6), the phonon Boltzmann equation now becomes

$$0 = -\frac{n_\lambda(\mathbf{q}) - n_\lambda^0(\mathbf{q})}{\tau_\lambda(\mathbf{q})} + e\nabla\varphi \cdot \sum_{\mathbf{k}\alpha,\mathbf{k}'\beta}\left[G_{\mathbf{k}\alpha}(\mathbf{k}'\beta,\mathbf{q}\lambda)\tau_{\mathbf{k}\alpha}\mathbf{v}_{\mathbf{k}\alpha}\frac{\partial f_{\mathbf{k}\alpha}^0}{\partial E} + G_{\mathbf{k}'\beta}(\mathbf{k}\alpha,\mathbf{q}\lambda)\tau_{\mathbf{k}'\beta}\mathbf{v}_{\mathbf{k}'\beta}\frac{\partial f_{\mathbf{k}'\beta}^0}{\partial E}\right]$$
(S9)

where the drift term vanishes because there is no temperature gradient. It can be readily solved to obtain the phonon distribution function

$$\Delta n_\lambda(\mathbf{q}) = \tau_\lambda(\mathbf{q}) e\nabla\varphi \cdot \sum_{\mathbf{k}\alpha,\mathbf{k}'\beta}\left[G_{\mathbf{k}\alpha}(\mathbf{k}'\beta,\mathbf{q}\lambda)\tau_{\mathbf{k}\alpha}\mathbf{v}_{\mathbf{k}\alpha}\frac{\partial f_{\mathbf{k}\alpha}^0}{\partial E} + G_{\mathbf{k}'\beta}(\mathbf{k}\alpha,\mathbf{q}\lambda)\tau_{\mathbf{k}'\beta}\mathbf{v}_{\mathbf{k}'\beta}\frac{\partial f_{\mathbf{k}'\beta}^0}{\partial E}\right]$$
(S10)

Considering the heat flow described by $\mathbf{q} = \sum_{\mathbf{q}\lambda}\hbar\omega_{\mathbf{q}\lambda}\mathbf{v}_{\mathbf{q}\lambda}\Delta n_{\mathbf{q}\lambda}$, we finally arrive at the phonon drag Peltier coefficient.

$$\Pi_{ph} = \frac{2e}{3\sigma\Omega N_{\mathbf{k}}N_{\mathbf{q}}k_BT}\sum_{\mathbf{q}\lambda}\hbar\omega_{\mathbf{q}\lambda}\tau_{\mathbf{q}\lambda}\mathbf{v}_{\mathbf{q}\lambda}\cdot\left[\sum_{\mathbf{k}\alpha,\mathbf{k}'\beta}\left(\tau_{\mathbf{k}\alpha}\mathbf{v}_{\mathbf{k}\alpha}-\tau_{\mathbf{k}'\beta}\mathbf{v}_{\mathbf{k}'\beta}\right)\cdot f^0_{\mathbf{k}\alpha}\left(1-f^0_{\mathbf{k}'\beta}\right)n^0_{\mathbf{q}\lambda}\cdot\frac{2\Pi}{\hbar}\right] \quad \text{(S11)}$$

$$\text{with} \quad \Pi = \pi\left|g_{\alpha\beta\lambda}(\mathbf{k},\mathbf{k}',\mathbf{q})\right|^2\cdot\delta\left(E_{\mathbf{k}'\beta}-E_{\mathbf{k}\alpha}-\hbar\omega_{\mathbf{q}\lambda}\right)\cdot\delta(\mathbf{k}'-\mathbf{k}-\mathbf{q})$$

The comparison of Eq. (S11) with Eq. (2) in the main text clearly shows the Kelvin relation ($\Pi_{ph} = TS_{ph}$).

**Supplementary Note 3. Simulation details**

Knowing all the quantities in Eq. (2) of the main text, phonon drag Seebeck coefficient can be calculated on an equal electron and phonon Brillouin zone mesh. The total electron relaxation time $\tau_{\mathbf{k}\alpha}$, as we have elaborated, combines both electron-phonon scattering as well as the electron-impurity scattering, according to Matthiessen's rule. Similarly, the phonon relaxation time $\tau_{\mathbf{q}\lambda}$ includes both phonon-phonon scattering, phonon-electron scattering and phonon-impurity scattering. Supplementary Table S1 shows the scattering mechanisms considered in this work and how they are treated.

**Supplementary Table S1.** Scattering mechanisms for electrons and phonons

| Carrier | Scattering | Method |
|---|---|---|
| Electron | Electron-phonon | DFT |
| | Electron-impurity | Brooks-Herring model(4, 8) |
| Phonon | Phonon-phonon | DFT |
| | Phonon-electron | DFT |
| | Phonon-impurity | Tamura model(9) |

The equilibrium properties of electrons and phonons are calculated from first principles using the QUANTUM ESPRESSO package(10) as described in Methods. Apart from these equilibrium properties, the key ingredients towards the first principles result are the quantities describing non-equilibrium properties. The electron phonon interaction matrix element $g_{\alpha\beta\lambda}(\mathbf{k},\mathbf{k}',\mathbf{q}) = \left(\frac{\hbar}{2m_0\omega_{\mathbf{q}\lambda}}\right)^{1/2} \cdot \langle \mathbf{k}'\beta | \partial_{\mathbf{q}\lambda} V | \mathbf{k}\alpha \rangle$ is one of them and leads to the electron relaxation time(11). The stringent convergence demands the knowledge of both wavefunctions and perturbing potentials on an ultra-dense mesh, which only became accessible recently due to a Wannier basis-based interpolation scheme(12), allowing us to interpolate between the matrix elements from the coarse meshs to produce finer meshes. Fine meshes up to 100×100×100 have been tested for their convergence (for both electrons and phonons). To obtain the electron relaxation time due to the electron-phonon coupling, we use the Gaussian smearing method. The convergence with respect to the Gaussian broadening parameter has also been checked. A band energy cutoff (measured from the band edge) is used to select only electron states near the band edge because states far away from the band edge will not contribute to the transport. Besides, the phonon relaxation times are also required. They are essentially related to the anharmonic force constants(13, 14), which can be obtained from first principles calculations. For heavily-doped samples, electron-phonon scattering of phonons also needs to be considered for calculating the phonon relaxation time. This is solved following our previous paper(7). Knowing all of these quantities, equation (2) in the main text (or Eq. (S11)) can be calculated at points on an equal electron and phonon Brillouin zone mesh, for which the tetrahedra integration method is implemented(15). To further speed up the calculation, based on the knowledge that only phonons that are close to the zone center contribute to the phonon drag effect, we define a wave vector cutoff, above which phonons will not be considered for equation (1). This cutoff has been checked and the change of the result is within 1% of the original value. Supplementary Table S2 lists some of the key parameters we used in this calculation.

Supplementary figure S1 shows the resulting intrinsic mobility and thermal conductivity with respect to temperature as a test to the electron and phonon relaxation times we obtain. We note here that previous first principles calculation have obtained similar agreement with experiments for the electron mobility(5, 6, 16) and the thermal conductivity(13, 14) in silicon. As is seen in supplementary figure S1, overall the results agree well with the experimental data. There is some discrepancy for the hole mobility in p-type silicon near room temperature. This shows that the experimental samples experience more scattering and therefore bear a lower mobility. We speculate this to be a result of the split valence bands due to the spin-orbit coupling at valence band edge. We cannot confirm this point yet because the spin-orbit coupling is not included in this work. However, this discrepancy should not affect the Seebeck coefficient calculation much, because we know from the Boltzmann description of the diffusive Seebeck coefficient and equation (1) that both of them will not be changed if the electron scattering time is just changed by a single constant factor. Therefore the results on p-type silicon can still give guidance in how to utilize the phonon drag effect in p-type materials.

For heavily-doped silicon, impurity scattering needs to be considered. Due to the lack of accurate and computationally feasible methods for calculating the impurity scattering, the effects from impurities are described using empirical models – the Brooks-Herring model for electron-impurity scattering(4, 8) and the Tamura model for phonon-impurity scattering(9). It was known that the Brooks-Herring model tends to underestimate the electron impurity scattering rate(4). We found from our calculation that the electron-impurity scattering has a small influence on the phonon drag effect. This small influence lies in the fact that the Seebeck coefficient essentially represents the ratio of the temperature-gradient induced current to the electric-field driven current. When the electron relaxation time is reduced, both of them are weakened and therefore the ratio between them is less affected. The Tamura model(9) is used to examine the effect of the ionized impurity scattering on the phonon drag effect, where the mass difference ratio $\Delta M / \bar{M}$ is chosen to be 1 to represent both the mass disorder and strain effect. We found from the calculation that phonon-ionized impurity scattering also only has small influences on the phonon drag effect. Based on the reasons given above, the use of empirical models in our calculation can be justified.

**Supplementary Table S2.** Parameters used in determining the electron and phonon relaxation times as well as in the calculation of the phonon drag effect. The parameter "$a$" in the last column is the lattice constant of silicon.

| **Quantities required** | Electron relaxation time | Phonon-phonon relaxation time | Electron-phonon scattering of phonons | Phonon drag effect |
|---|---|---|---|---|
| **k**-mesh (electron) | $70^3 \sim 100^3$ |  | $70^3$ | $70^3 \sim 100^3$ |
| **q**-mesh (phonon) | $80^3$ | $70^3 \sim 100^3$ | $70^3$ | $70^3 \sim 100^3$ |
| Integration method (broadening parameter if any) | Gaussian (0.002eV) | Gaussian (1cm$^{-1}$) | Tetrahedra | Tetrahedra |
| Energy / wavevector (**q**) cutoff | 0.5 eV | $\mathbf{q}:(0.2 \sim 2)\dfrac{2\pi}{a}$ | 1.0 eV | energy: 0.5eV $\mathbf{q}:(0.2 \sim 2)\dfrac{2\pi}{a}$ |
| Nearest neighbor considered in the force constant fitting(13) |  | 2$^{nd}$ force constant (harmonic): 7  3$^{rd}$ force constant (anharmonic): 1 |  |  |

## Supplementary Note 4. Electron mode contribution to phonon drag

Here we show the mode-specific contributions to the phonon drag Seebeck coefficient as well as the electrical conductivity and diffusive Seebeck coefficient from the electron side. This provides us the knowledge of what portion of electrons contributes to these transport properties and especially the phonon drag effect most notably, which will be used when we discuss the usage of nanoclusters to selectively scatter phonons. Equation (2) of the main text directly presents the phonon mode-specific contribution to the phonon drag (term inside the bracket). For the electrons, we can similarly combine all the terms that are labeled with the same electron wave vector **k** and band number $\alpha$ in the summation. The result is given in supplementary figure S4 for the lightly-doped n-type silicon with a doping concentration of $10^{14}$ cm$^{-3}$.

In general, we see that at the same temperature the accumulated contribution curves for the three physical quantities (electrical conductivity, diffusive Seebeck coefficient and phonon drag Seebeck coefficient) almost overlap. This is because the Fermi-Dirac distribution, which modifies the population of the electrons, changes more strongly with the electron states compared to other properties such as scattering rates, and essentially confines the electron states that are important for transport properties to a small region near the band edge. Therefore we see a general monotonically-increasing accumulated contribution curve. To be more specific, we see in supplementary figure S4a that, when temperature decreases, the curves move towards the band edge. This is a result of the temperature characteristics of Fermi-Dirac distribution function, which decreases more rapidly with energy as temperature decreases. As a result, the electrons that participate in the transport are more confined to the band edge at lower temperatures. At 300K, most electron states that are important for the transport properties are located within 0.2eV from the band edge. Because electrons are confined to the band edge, which is made up of six equivalent electron pockets, and the conduction band minimum corresponds to a wavelength of 0.67nm, significant contributions to the transport properties should come from electrons with wavelength around 0.67nm. In supplementary figure S4b, indeed we see that most of the contributions come from electrons with wavelength between 0.6nm and 0.7nm. At lower temperatures, the curve becomes slightly narrower, and the reason is the same as before: electrons become more confined to the band edge and the reciprocal space they occupy then shrinks. In terms of the electron mean free path, Supplementary figure S4c indicates that, electrons at 300K have mean free paths between 20nm and 80nm. The mean free path increases as temperature decreases, and at 100K the majority of electrons have mean free paths around 100nm~300nm. We should note that these plots are obtained for a lightly-doped silicon. For the heavily-doped silicon, the characteristics of the energy-dependence(5) and wavelength-dependence will remain the same, meaning that electrons involved in the transport process are still confined within ~0.2eV from the band edge and have wavelengths around 0.6nm~0.7nm, because the qualitative argument given above does not change. In comparison, mean free paths of electrons will decrease due to the impurity scattering, and therefore the mean free path accumulated curve will move towards the left. For example, it was known that for the n-type silicon with $10^{19}$ cm$^{-3}$ doping concentration, the electrons have mean free paths below 20nm within the temperature range of 100K~300K(5).

## Supplementary Note 5. Nanocluster scattering as a phonon frequency filter

Here we examine the nanocluster scattering to effectively scatter high-frequency phonons. Nanoclusters are clusters that have impurity atoms different from the host atoms with sizes ranging from sub-nanometer to a few nanometers. One extreme case of the nanocluster is a single impurity atom embedded in the host, for which the theoretical model developed by Tamura can be used to estimate the phonon-impurity scattering(9). For clusters that contain more than one impurity atom, there has been development of first principles approach based on Green's function calculation(17, 18), which can provide more accurate results. For simplicity, we will not use such a rigorous method to describe the nanocluster scattering. Instead, we use an analytical formula(17) generalized from the Tamura model for the description of the nanocluster scattering effect. In the Born approximation, it can be shown that the phonon-nanocluster scattering rate is

$$\tau_{imp}^{-1}(\mathbf{q},\lambda) = \frac{\pi}{12N} f \left(\frac{\Delta M}{\bar{M}}\right)^2 \omega_{\mathbf{q}\lambda}^2 \cdot D^*(\omega_{\mathbf{q}\lambda})$$

$$D^*(\omega_{\mathbf{q}\lambda}) = 6 \sum_{\mathbf{q}'\lambda'} \left[ \left| \sum_\sigma \vec{e}_\sigma^*(\mathbf{q}',\lambda') \cdot \vec{e}_\sigma(\mathbf{q},\lambda) \right|^2 \cdot |S_{\Delta\mathbf{q}}|^2 \cdot \delta(\omega_{\mathbf{q}\lambda} - \omega_{\mathbf{q}'\lambda'}) \right]$$

(S12)

where $N$ is the total number of unit cells, $f$ is the volume fraction of the nanoclusters, $\Delta M$ is the mass difference of the impurity atom and the host atom, $\bar{M}$ is the average mass of all the atoms, $\omega_{\mathbf{q}\lambda}$ describes the phonon frequency and $\vec{e}_\sigma(\mathbf{q},\lambda)$ is the unit vector along the polarization of the atom labeled by $\sigma$ in the unit cell. $S_{\Delta\mathbf{q}}$ is the structure factor with the sum includes all the unit cells occupied by one nanocluster. $D^*(\omega_{\mathbf{q}\lambda})$ can be regarded as a generalized phonon density of state. It can be shown that when the nanocluster contains only one impurity atom (the structure factor is one in this case), $D^*(\omega_{\mathbf{q}\lambda})$ reduces to the normal phonon density of state, and as a result, Eq. (S12) is essentially the same as the Tamura model. The unit vectors along the polarizations of the atoms as well as the phonon frequencies are obtained from first principle calculations. The mass fraction term $\frac{\Delta M}{\overline{M}}$ is chosen to be 1 to represent both mass disorder and force constant disorder. This is a typical number for alloys. For example, if the host is silicon, then the addition of germanium atoms act as impurities with a mass fraction of around 1.6.

We test nanocluster size up to 1nm (this is the equivalent diameter defined through the total volume of the unit cells contained in the nanocluster). For the value of 1nm, electron wavelengths are comparable to the nanoclusters size (see supplementary figure S4). We can use the geometric limit to estimate the upper bound for the electron-nanocluster scattering, which gives a corresponding mean free path of $\Lambda_{nanoparticle} = \frac{4r}{3f} \approx 330\text{nm}$, where $r$ is the characteristic radius of the nanoparticle and chosen to be 0.5nm for estimation (volume fraction $f$ here is chosen to be 0.2%, which is the maximum value we set). For nanoclusters that are smaller, the geometric limit becomes smaller. However, the wavelengths are now large compared to the nanocluster sizes and enter into the Rayleigh scattering regime, where the scattering rate falls below the geometric limit. Therefore $\Lambda_{nanoparticle}$ should be on the order of 300nm or even larger. From previous work on first principles calculation of silicon(5) we know that for the doping concentration of $10^{19}$ cm$^{-3}$ the electron mean free paths are less than 20nm. The comparison with $\Lambda_{nanoparticle}$ indicates that the dominant scatterings for electrons still come from the phonons and dopants. Therefore the nanocluster scattering for electrons can be neglected and the electrical conductivity is barely affected.

**Supplementary Figure Legends**

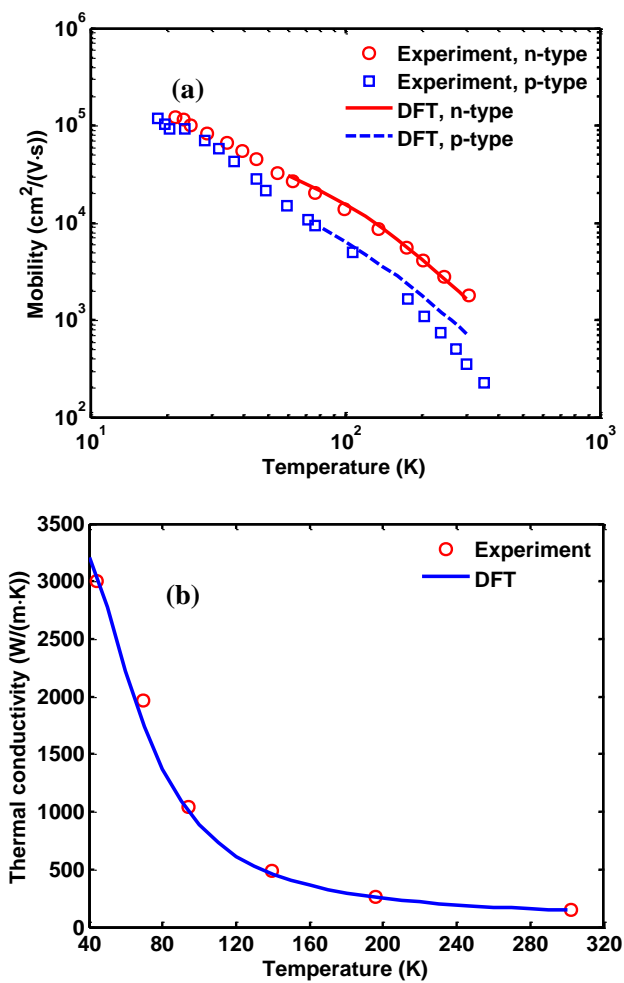

**Supplementary Figure S1.** (a) Temperature dependence of the intrinsic mobility in n-type and p-type silicon compared with that of sufficiently pure samples(19) as well as (b) the thermal conductivity of pure silicon compared with the experiments(20). The intrinsic mobility is calculated assuming a carrier concentration of $10^{14}$ cm$^{-3}$.

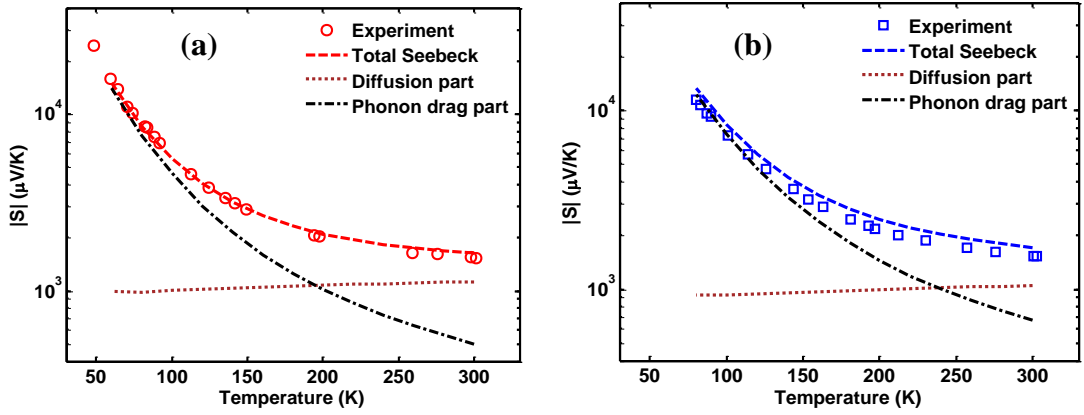

**Supplementary Figure S2**. Phonon drag Seebeck coefficient for (a) electrons and (b) holes in lightly-doped silicon. The open circles and squares are taken from the experiment(21), with the corresponding net doping concentration of $2.8 \times 10^{14} \text{cm}^{-3}$ for electrons and $8.1 \times 10^{14} \text{cm}^{-3}$ for holes, respectively. Lines are first principles results assuming the same doping concentrations. Dotted lines represent the diffusive Seebeck coefficient while dash-and-dot lines represent the phonon drag contribution with respect to the temperature on a semilog plot. The phonon drag contribution increases dramatically as the temperature decreases and converges to the total Seebeck coefficient, shown by the dashed lines. The experimental data(21) with larger sample size and lower net doping concentration are chosen as comparison.

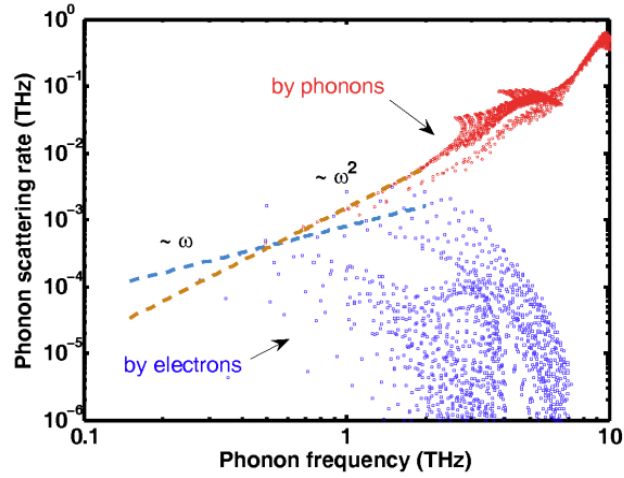

**Supplementary Figure S3.** Phonon scattering rates by phonons themselves and by electrons. The phonon-phonon scattering rate follows a $1/\tau \sim \omega^2$ trend for low-frequency phonons. The phonon-electron scattering only roughly follows a $1/\tau \sim \omega$ because this scaling behavior is valid only for very low frequency phonons and does not consider the anisotropy of the energy bands(7). Nonetheless, it is clearly seen that below around 1THz, the phonon-electron scattering starts to dominate over the intrinsic phonon-phonon scattering. The calculation is carried out for the n-type silicon with a $10^{19}$ cm$^{-3}$ doping concentration at 300K.

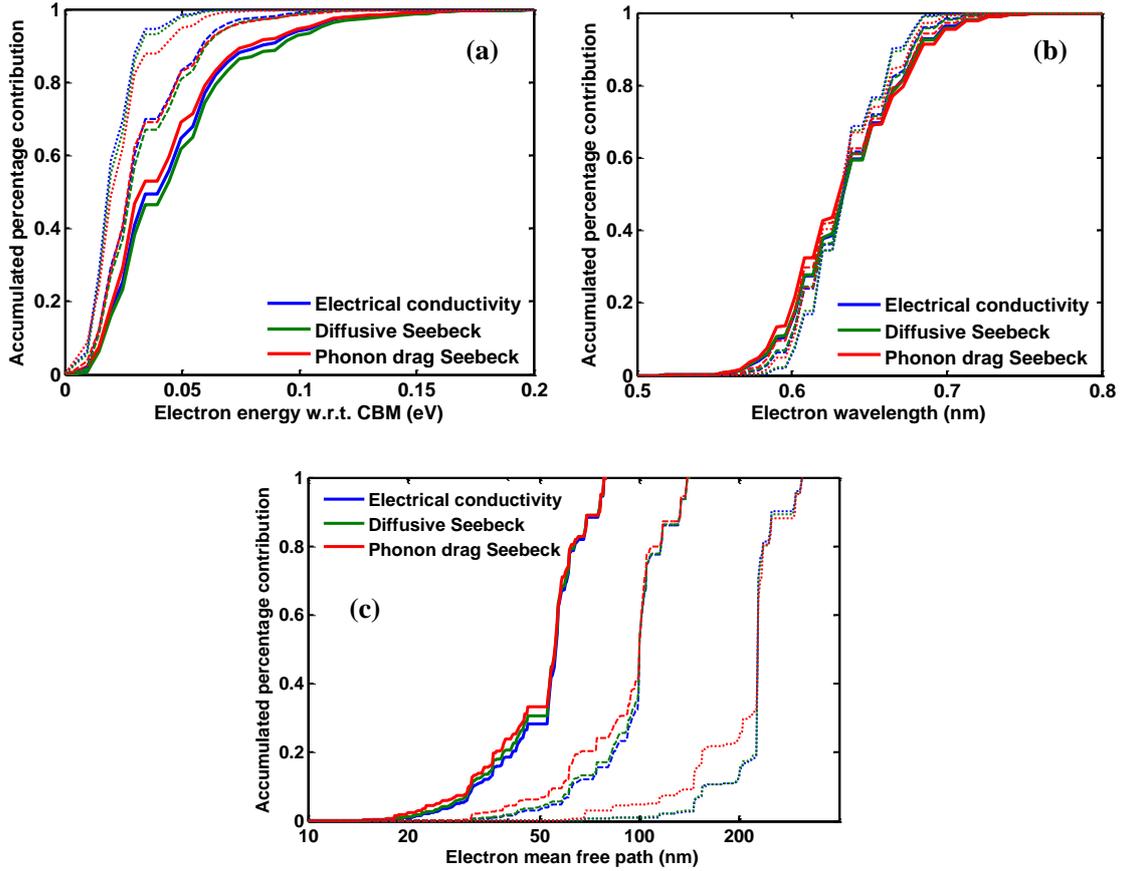

**Supplementary Figure S4.** Accumulated contribution to electrical conductivity, diffusive Seebeck coefficient and phonon drag Seebeck coefficient, with respect to (a) electron energy (measured from the conduction band edge), (b) electron wavelength and (c) electron mean free path. Physical quantities are labeled with different colors (blue for electrical conductivity, green for diffusive Seebeck coefficient and red for phonon drag Seebeck coefficient). In all three plots, solid curves describe results at 300K, while dashed curves represent 200K and dotted lines 100K. The results are calculated assuming a doping concentration of $10^{14}$ cm$^{-3}$.

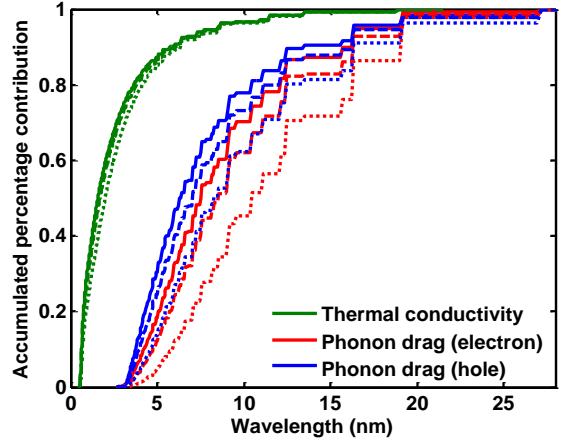

**Supplementary Figure S5.** Phonon mode-specific accumulated contributions to the phonon drag Seebeck coefficient and the thermal conductivity with respect to the phonon wavelength. Solid lines show the contribution at 300K, while dashed lines are used at 200K and dotted lines at 100K. Green curves show results for the thermal conductivity. Red and blue curves represent results for phonon drag in n-type and p-type silicon, respectively. Clearly phonons that are significant in phonon drag typically have longer wavelengths. Besides, we can see that the spectral difference between the contributions of the phonon modes to the phonon drag Seebeck coefficient and to the thermal conductivity becomes larger at lower temperatures.

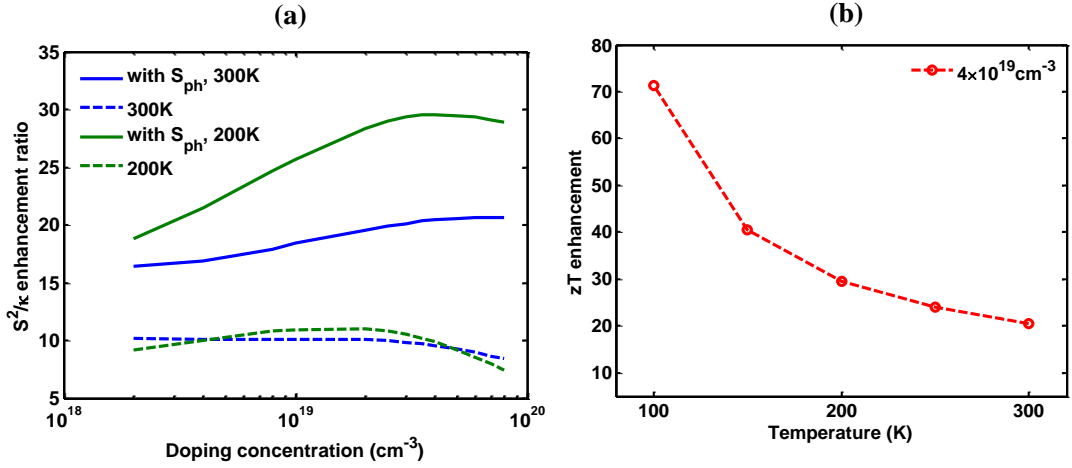

**Supplementary Figure S6.** (a) The enhancement of the factor $S^2/\kappa$ as a function of doping concentration when phonon modes are selectively scattered and (b) the resulting zT enhancement assuming electrical conductivity is not affected at a doping concentration of $4\times10^{19}\,\text{cm}^{-3}$ in n-type silicon. In (a), solid lines represent the results where phonon drag is included and preferable phonon modes are chosen, while for the dashed lines it is assumed that phonon drag is neglected. The results examined in (a) assume that the thermal conductivity is reduced to $4\,\text{W}/(\text{m}\cdot\text{K})$ for all the curves.

Without the phonon drag effect, the enhancement of $S^2/\kappa$ is around 10, which barely changes from 300K to 200K, because the diffusion contribution to the Seebeck coefficient decreases while the thermal conductivity actually increases as the temperature decreases, which tend to cancel each other. The situation is different if the phonon drag effect is included. First, because the phonon drag magnitude is comparable to the diffusion contribution, the inclusion of phonon drag with selective phonon modes helps to boost $S^2/\kappa$. Furthermore, when the temperature is lowered, the phonon drag effect becomes even more pronounced and therefore makes the enhancement of $S^2/\kappa$ even larger. The latter point is more clearly seen in (b), where we fix the doping concentration to be $4\times10^{19}\,\text{cm}^{-3}$ and vary the temperature. The enhancement of $S^2/\kappa$ directly translates to the zT enhancement compared to the bulk doped silicon at same temperatures, reaching a value as large as ~70 at 100K.

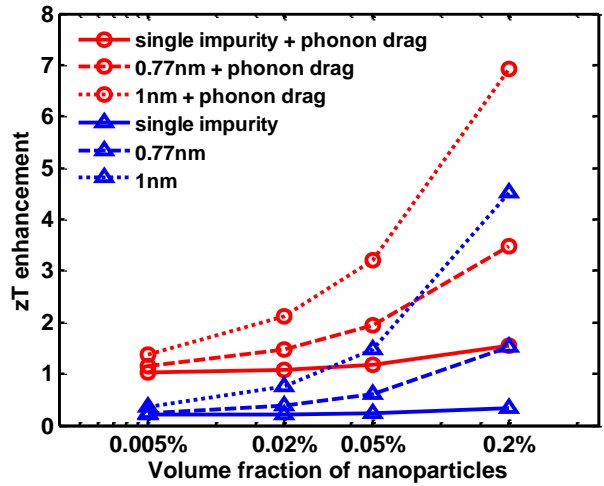

**Supplementary Figure S7.** The enhancement of the thermoelectric figure of merit zT as volume fraction of nanoclusters using phonon frequency selectivity. The calculation is done for the n-type silicon with the doping concentration of $10^{19}$ cm$^{-3}$ at 200K. Curves labeled with phonon drag include the phonon drag contribution to the Seebeck coefficient while others neglect the phonon drag effect. It can be seen that, using 1nm size nanoclusters with a volume fraction of 0.2% at 200K, the enhancement can reach above 7, and the neglect of the phonon drag effect will reduce the Seebeck coefficient and therefore the enhancement as well. At the same volume fraction, nanoclusters with larger size scatter phonons more strongly. This fact is known for decades (2) and is utilized here to select low-frequency phonons. However, nanoclusters that are much larger will also significantly reduce the phonon drag effect. Therefore there will be an optimal choice for the size of the nanoclusters and our calculations show that this optimal size is around 1nm for silicon at a doping concentration of $10^{19}$ cm$^{-3}$.